\documentclass[10pt]{article}
\usepackage{amssymb}
\usepackage{latexsym}
\usepackage[dvips]{epsfig}

\newtheorem{proposition}{Proposition}
\newtheorem{lemma}{Lemma}

\def\ethbar{\overline{\eth}}
\def\gammab{\overline{\gamma}}
\def\sigmab{\overline{\sigma}}
\def\nub{\overline{\nu}}
\def\cbar{\overline{c}}
\def\lambdab{\overline{\lambda}}
\def\pib{\overline{\pi}}

\def\alphab{\overline{\alpha}}
\def\mbar{\overline{m}}
\def\xib{\overline{\xi}}
\def\deltab{\overline{\delta}}
\def\betab{\overline{\beta}}
\def\taub{\overline{\tau}}
\def\zetab{\overline{\zeta}}

\def\Re{\mbox{Re}}
\def\Im{\mbox{Im}}
\def\O{\mbox{O}}
\def\N{\mathcal{N}}
\def\Z{\mathcal{Z}}
\def\d{\mbox{d}}

\font\SYM=msbm10
\newcommand{\Real}{{\SYM R}}
\newcommand{\Complex}{{\SYM C}}

\font\tenscr=rsfs10 scaled1100
\font\sevenscr=rsfs7 
\font\fivescr=rsfs5 
\skewchar\tenscr='177 \skewchar\sevenscr='177 \skewchar\fivescr='177
\newfam\scrfam
\textfont\scrfam=\tenscr
\scriptfont\scrfam=\sevenscr
\scriptscriptfont\scrfam=\fivescr

\def\scri{{\fam\scrfam I}}

\begin{document}

\title{On Killing vector fields and Newman-Penrose constants.}
\author{Juan Antonio Valiente Kroon \thanks{E-mail address: {\tt j.a.valiente@qmw.ac.uk}} \\
School of Mathematical Sciences,\\
Queen Mary \& Westfield College,\\
Mile End Road, London E1 4NS,\\
United Kingdom.}

\maketitle

\begin{abstract}
Asymptotically flat spacetimes with one Killing vector field are studied. The Killing equations are solved asymptotically using polyhomogeneous expansions (i.e. series in powers of $1/r$ and $\ln r$), and solved order by order. The solution to the leading terms of these expansions yield the the asymptotic form of the Killing vector field. The possible classes of Killing fields are discussed by analysing their orbits on null infinity. The integrability conditions of the Killing equations are used to obtain constraints on the components of the Weyl tensor ($\Psi_0$, $\Psi_1$, $\Psi_2$) and on the shear ($\sigma$). The behaviour of the solutions to the constraint equations is studied. It is shown that for Killing fields that are non-supertranslational the characteristics of the constraint equations are the orbits of the restriction of the Killing field to null infinity. As an application, the particular case of boost-rotation symmetric spacetimes is considered. The constraints on $\Psi_0$ are used to study the behaviour of the coefficients that give rise to the Newman-Penrose constants, if the spacetime is non-polyhomogeneous, or the logarithmic Newman-Penrose constants, if the spacetime is polyhomogeneous.
\end{abstract}

\section{Introduction}
The Newman-Penrose (NP) constants \cite{np68} are a set of five complex quantities, defined for asymptotically flat spacetimes with a smooth null infinity, with the remarkable property of being absolutely conserved even in the presence of gravitational radiation. Recently \cite{javk98a}, \cite{javk98c}, it has been shown that in  a more general setting ---that of polyhomogeneous spacetimes--- the NP constants are not conserved, but nevertheless, an adequate generalization of them (logarithmic Newman-Penrose constants) can be constructed which are indeed conserved. Polyhomogeneous spacetimes are spacetimes which are expanded asymptotically in terms of combinations of powers of $1/r$ and $\ln r$. The introduction of this more general kind of expansion carries a drawback: null infinity ($\scri$) is no longer smooth. For more details on these, and other aspects of polyhomogeneity, we refer the reader to references \cite{xiv}, \cite{javk98a}, \cite{javk98b}, and \cite{javk98c}.

The physical meaning of the NP and logarithmic NP constants is still an open question. The work of interpretation has not been easy for a number of reasons. One of them is a lack of examples of exact solutions to the field equations representing physically sensible radiative asymptotically flat spacetimes. No explicit radiative solution which satisfies all the requirements of asymptotic flatness is known. The examples one can make use of reduce essentially to the family of boost-rotation symmetric spacetimes \cite{bicak0}, \cite{bicak1}, \cite{bicak2}. These spacetimes have two commuting Killing vectors, one of them axial, and the other one, such that it leaves invariant the origin's null cone. This family includes, among others, the Bonnor-Swaminarayan \cite{bonnor} and the C metric. Bi\v{c}\'{a}k, Hoenselaers \& Schmidt have given a systematic method of constructing these spacetimes \cite{bicak4}, \cite{bicak5}.

Boost rotation-symmetric spacetimes  describe ``uniformly accelerated particles'' that  approach the speed of light asymptotically. The smoothness of the solution requires the spacetimes to be reflection symmetric \cite{bicak2}; therefore at least two particles with opposite acceleration are present, and future null infinity contains at least two singular points. The null infinity of a boost-rotation symmetric spacetime can be global, in the sense that it admits spherical cuts, but the generators are not complete. Ashtekar \& Dray \cite{ashtekar} have shown that the C-metric admits a conformal completion such that the cuts of $\scri$ are the 2-sphere $S^2$. This example in particular settled the issue of the existence of radiative asymptotically flat solutions to the Einstein-Maxwell field equations. 

 The boost-rotation symmetric spacetimes are usually given in a coordinate system that clearly exhibits their symmetries. The transformation between this coordinate system and the Bondi coordinates used in the asymptotic expansions of the gravitational field  has to be given in terms of series. These expansions are extremely messy, and usually only the leading terms can be calculated explicitly. To add to the problem, the coefficients in terms of which the NP and logarithmic NP constants are defined are found deep into the series expansions. As a curious observation, Kinnersley \& Walker \cite{kinnersley} in a note added in proof mention that the C metric is a counterexample to the claim that all algebraically special spacetimes have zero NP constants. However, no expression for the conserved quantities is given.

Bi\v{c}\'{a}k \& Schmidt \cite{bicak}, starting with a vacuum axially symmetric spacetime, imposed an extra symmetry on the spacetime. Expanding the Killing equations in powers of $1/r$, and solving order by order one finds an asymptotic expression for the new Killing field. The boost-rotation symmetry appears playing a privileged role as the only other symmetry an asymptotically flat axially symmetric radiative spacetime can have. A generalization of this work for the case of electro-vacuum spacetime has been  done recently  by Bi\v{c}\'{a}k \& Pravdov\'{a} \cite{bicak3}.

It is clear that if one  wants to pose an initial value problem for a spacetime that has a particular symmetry, the initial data cannot be arbitrary, for it should satisfy some constraints imposed by the Killing vector field. For instance, if the spacetime is bound to be axisymmetric, the initial data cannot depend on $\varphi$. In the analysis by Bi\v{c}\'{a}k \& Schmidt, these constraint equations begin to appear when the expansions are carried to higher orders in $1/r$. In reference \cite{bicak}, a constraint for the news function was found, and in \cite{bicak3} another for the mass aspect.

In this article their approach will be generalized in two ways: first it will not be assumed from the beginning that the spacetime is axially symmetric, and second it will be assumed that the spacetimes can be expanded asymptotically using polyhomogeneous series. The Killing equations will be solved using these expansions, and the constraints for the different quantities of interest will be deduced from the integrability conditions of the Killing equations. In particular, we will be most interested in the constraint on the coefficient of order 6 in $1/r$ of $\Psi_0$, as it is the one from which the Newman-Penrose (and the logarithmic NP) constants are calculated.

As it will be shown, the structure of the constraint equations is very similar to that of the continuity equations of the classical mechanics of continuos media. This fact will allow us to gain some insight on the behaviour of the solutions to the constraint equations, and put forward a tentative interpretation of the physical mechanisms involved, although it may not be possible to calculate the explicit form of the solutions.

The article is organized as follows: in section 2, some preliminaries are discussed. These include a brief note on the coordinate system and the null tetrad to be used, the Killing equations and their integrability conditions in the NP formalism, some comments on the hypothesis of polyhomogeneity, and some remarks on the characteristic initial value that will prove of use on later discussions. In section 3 the Killing equations are solved to the first order yielding the asymptotic form of the Killing fields compatible with asymptotic flatness. The case of axial symmetry is analysed. The result by Bi\v{c}\'ak \& Schmidt \cite{bicak} on the privileged role of the boost-rotation symmetry in axisymmetric spacetimes is recovered. Section 4 is devoted to the study of the constraint equations that can be deduced from the integrability conditions of the Killing equations. The general form of these constraints is discussed, and some general remarks on the behaviour of their solutions are put forward. Some emphasis is put on the characteristic curves of the differential equations. The constraints for the leading term of the shear ($\sigma_{2,0}$), the news function ($\dot{\sigma}_{2,0}$), the mass aspect ($\mbox{Re}\, \Psi_2^{3,0}$) and the coefficients $\Psi_0^4$, $\Psi_0^5$, $\Psi_0^6$ are calculated and analysed. The resulting expressions are applied to the particular case of boost-rotation symmetric spacetimes. The final product of this analysis is the constraint equation for the coefficient $\Psi_0^{6,X}$ that gives rise to the NP constants. From these constraints, some remarks on the interpretation of these quantities are made.

There are 4 appendices. Appendices A and B discuss the solutions to two partial differential equations that will be used in the body of the article. In appendix C, the orbits of the restriction of the Killing field to null infinity are studied, and a classification of them is done. Finally in appendix D some spin-weighted spherical harmonics are listed for quick reference.

\section{Preliminaries}

\subsection{Coordinates \& tetrad}
Most of the calculations will be done with the NP formalism. The coordinates and tetrad used are the same as the ones described in Stewart's book \cite{stewart}. The coordinates $(x^0,x^1,x^2,x^3)=(u,r,\theta,\varphi)$ are such that $u$ is a \emph{retarded time} labelling the foliation of the spacetime $\mathcal{M}$ by null hypersurfaces. The \emph{affine parameter} $r$ parametrises the geodesic generators of the null hypersurfaces, and the angular coordinates $(\theta,\varphi)$ are such that they are constants along the generators of $\scri$ and along the geodesic generators of the null hypersurfaces. The freedom left in this coordinate construction is: a relabelling of the null hypersurfaces, a different choice of angular coordinates, and a rescaling and setting of the origin for the affine parameter $r$.

The tetrad is constructed so that $l^\mu$ is tangent to the geodesic generators of the null hypersurfaces; $n^\mu$ is future pointing, and orthogonal to the 2-surfaces $u=const.$, $r=const.$ ($S_{u,r}$); and $m^\mu$, and $\mbar^\mu$ span the tangent space to $S_{u,r}$ ($\mbox{T}(S_{u,r})$). The freedom left in this tetrad construction consists in: a boost $l^\mu \mapsto Al^\mu$, $n^\mu \mapsto A^{-1}n^\mu$ which yields a rescaling of $r$; and a spin $m^\mu \mapsto e^{i\vartheta}m^\mu$. 

It can be shown that 
\begin{eqnarray}
l^\mu&=& \delta^\mu_1, \\
n^\mu&=& \delta^\mu_0 + Q \delta^\mu_1 + C^i\delta^\mu_i, \\
m^\mu&=& \xi^i \delta^\mu_i.
\end{eqnarray}
This tetrad is such that  $\epsilon =\kappa =0$, $\tau =\overline{\pi }=\overline{\alpha }+\beta $, and $\rho$ and $\mu$ are real functions.

The freedom left in the construction of the vectors $m^\mu$ and $\mbar^\mu$ gives rise to the notion of \emph{spin-weighted quantities}. The derivatives $\eth$ and $\ethbar$ work as raising and lowering for the spin weight of the relevant quantities. We will stick to Penrose \& Rindler's convention \cite{penroserindleri}. The connection between the $\eth$ and $\ethbar$ operators and the directional derivatives $\delta$ and $\deltab$ is given by
\begin{eqnarray}
\eth \eta&=& \delta \eta +s(\alphab-\beta)\eta, \\
\ethbar \eta &=& \deltab \eta -s(\alpha -\betab) \eta,
\end{eqnarray}
where $\eta$ is a quantity of spin weight $s$. A spin-weighted quantity on the sphere ($S^2$) can be expanded in terms of \emph{spin-weighted spherical harmonics}. These spherical harmonics are a complete and orthonormal set of eigenfunctions of the operator $\ethbar\eth$. 

When dealing with differential equations that contain the operators $\eth$ and $\ethbar$, the following lemma will be most useful \cite{stewart}.
\begin{lemma}
 Suppose $\eta$ is continuous on the sphere and has spin weight $s>0$. Then if $ \ethbar\eta=0$ it follows that $\eta=0$, and if $\eth \eta=0$ then $\eta$ is a linear combination of the ${_sY_{s,m}}$. The analogous results hold for $s<0$ by interchanging $\eth$ and $\ethbar$.
\end{lemma}

\subsection{Polyhomogeneity}
In order to handle with ease polyhomogeneous expansions, some conventions will be adopted. We will refer to the Newman-Penrose field equations, the Bianchi identities and the frame equations in the way they are labeled in Stewart's book ((a) to (r), (Ba) to (Bk), and (Fa) to (Ff)). The Killing equations, (\ref{K1}) to (\ref{K7}) will be referred as (K1) to (K7). A first subscript in the label of a given equation will mean that we are just interested in that particular power of $1/r$, and a second subindex will refer to a particular power of $z=\ln r$.

 Otherwise stated, it will be assumed that all quantities (components of the Weyl tensor, spin coefficients, tetrad functions, and coefficients of the Killing fields) are polyhomogeneous functions. We will say that a function is polyhomogeneous if in a neighborhood of $\scri$ it can be written as
\begin{equation}
f=\sum^\infty_{k=1} f_k r^{-k} = \sum^\infty_{k=1} \sum_{j=0}^{N_k} f_{kj} r^{-k} \ln^j r,
\end{equation}
where $f_k$ is a polynomial of degree $N_k$ in $z=\ln r$, whereas $f_{kj}$ is a function of $(u,\theta,\phi)$  ---no dependence on $r$ left---. Let $\#$ denote the degree of a polynomial. In the example given above, $\#f_k=N_k$. Sometimes, the degree of the polynomial will appear next to it in square brackets, $f_k[N_k]$.

The $\eth$ operator has a polyhomogeneous expansion
\begin{equation}
\eth=r^{-1}\eth_1[0] +r^{-2}\eth_2[N_3+1]+\dots,
\end{equation}
where 
\begin{equation}
\eth_1\eta(\theta,\varphi)=\delta_1\eta(\theta,\varphi)+2s\alphab_1\eta(\theta,\varphi),
\end{equation}
and $\eth_1 f(r)=0$.

\subsection{The asymptotic characteristic initial value problem}

The asymptotic characteristic initial value problem is usually set by supplying $\Psi_0$ on an initial null hypersurface $\N_0$, $\sigma_{2,0}$ on $\scri^+$ (or $\scri^-$), and $\Psi_1^{4,0}$, $\Psi_2^{2,0}$, $\xi^i_1$ on $\Z_0=\scri^+ \cap \N_0$ (or $\Z_0=\scri^- \cap \N_0$). K\'{a}nn\'{a}r \cite{kannar} has proven the well posedness and existence/uniqueness of the initial value problem for data that is $C^\infty$. A similar theorem for polyhomogeneous initial data is not yet available. However, one can use K\'{a}nn\'{a}r's result as a sensible guide when formally solving the initial value problem with polyhomogeneous data.

The most general form for the component $\Psi_0$ of the Weyl tensor \cite{javk98a} is given by
\begin{equation}
\Psi_0=\O(r^{-3}\ln^{N_3} r),
\end{equation}
that is
\begin{equation}
\Psi_0= \Psi_0^3[N_3]r^{-3} + \Psi_0^4[N_4]r^{-4} + \Psi_0^5[N_5]r^{-5}+\Psi_0^6[N_6]r^{-6}+ \cdots.
\end{equation}
However, in order to ease the calculations, we will restrict our attention to spacetimes such that

\begin{equation}
\Psi_0=\O(r^{-4}\ln^{N_4} r), 
\end{equation}
This family of polyhomogeneous spacetimes are such that the leading term of the shear ($\sigma_2$) contains no logarithmic terms (finite shear at $\scri$). Among this family, one finds the ``minimal'' polyhomogeneous spacetimes, which are those whose logarithmic terms are directly due to non-compliance with the \emph{Outgoing Radiation Condition}. One has reasons to suspect that these particular group of polyhomogeneous spacetimes are the ones with physical relevance \cite{xiv},\cite{javk98b}. The component $\Psi_0$ for these ``minimal'' spacetimes has the form

\begin{equation}
\Psi_0=\Psi_0^4 r^{-4} + \left( \Psi_0^{5,0} +\Psi_0^{5,1} \ln r \right)r^{-5}+\left( \Psi_0^{6,0} +\Psi_0^{6,1} \ln r \right)r^{-6}+ \cdots,
\end{equation}
while the non-polyhomogeneous spacetimes obey the Peeling theorem:

\begin{equation}
\Psi_0= \Psi_0^{5,0}r^{-5} + \Psi_0^{6,0}r^{-6}+ \cdots.
\end{equation}

Details on how to solve the NP hierarchy using polyhomogeneous expansions can be found in references \cite{javk98a} and \cite{javk98c}. The results given there will be used whenever it is necessary.

The logarithmic Newman-Penrose constants are given in terms of $\Psi_0$ by
\begin{equation}
\mathcal{Q}_k^X=\int_{S^2}\Psi _0^{6,X}(_2\overline{Y}_{2k})\mbox{d}S, \label{logconst}
\end{equation}
where $X=\max \{N_{5},3N_3+3,N_3+N_4+2\}$ \cite{javk98c}. In the case of non-polyhomogeneous spacetimes ($N_3=N_4=-\infty$, and $N_5=0$ so that $X=0$) one recovers the original NP constants.

\subsection{The Killing equations and their integrability conditions in the NP formalism}

The Killing equations and their integrability conditions ($\mathcal{L}_{\xi^\mu}C_{\nu\mu\lambda\chi}=0$, where $C_{\nu\mu\lambda\chi}$ is the Weyl tensor) written in terms of GHP quantities can be found for example in the article by Kolassis \& Ludwig \cite{kolassis} (a NP version had been worked out previously by Collison \& French \cite{collinson}, but it is known to have several typographical errors). The equations given in the present article were deduced from theirs, and particularized to the specific NP null tetrad that was described in a previous section.

Let $\xi ^\mu $ be a Killing vector. It can be written in terms of the
vectors of the  null tetrad as

\begin{equation}
\xi ^\mu =al^\mu +bn^\mu -cm^\mu -\cbar \,\overline{m}^\mu,
\end{equation}
where $a$, $b$\  $\epsilon$ \ \Real \  have spin-weight $0$, and $c$\ $\epsilon$\ \Complex \  has spin-weight $-1$. The Killing equations (equations (K1) to (K7)) for the tetrad used in this article are:
\begin{eqnarray}
Db &=&0, \label{K1} \\
\Delta a+(\gamma +\gammab )a &=&\nub c+\nu \cbar, \label{K2}\\
\ethbar c &=&-\sigmab a+\lambda b, \label{K3} \\
\Delta b-(\gamma +\gammab )b+Da &=&0, \label{K4} \\
\eth c+\ethbar \cbar &=&-2\rho a+2\mu b, \label{K5} \\
Dc+\ethbar b &=&\pi b-\rho c-\sigmab \, \cbar, \label{K6}\\
\Delta \cbar-(\gamma -\gammab )\cbar+\eth a &=&-\tau
a+\lambdab c+\nub b+\mu \cbar \label{K7},
\end{eqnarray}
and the integrability conditions for $\sigma$, $\Psi_0$, $\Psi_1$ and $\Psi_3$ (equations (IS), (I0), (I1), (I2)) are:
\begin{equation}
aD\sigmab+b\Delta\sigmab -c\eth\sigmab -\cbar\ethbar\sigmab +b\sigmab \, \gamma-3b\sigmab \, \gammab=\ethbar\mathcal{Q}-\taub\mathcal{Q} + (\mathcal{P}+2i\mathcal{S})\sigmab, \label{IS}
\end{equation}
\begin{equation}
aD\Psi_0 + b\Delta\Psi_0 -c\eth\Psi_0 -\cbar\ethbar\Psi_0 -4b \gamma\Psi_0= 2(\mathcal{P}-i\mathcal{S})\Psi_0 -4 \overline{\mathcal{Q}}\Psi_1, \label{I0}
\end{equation} 
\begin{equation}
aD\Psi_1 +b\Delta \Psi_1 -c\eth\Psi_1 -\cbar\eth\Psi_1 -2b\gamma\Psi_1=(\mathcal{P}-i\mathcal{S})\Psi_1 - \overline{\mathcal{Q}}^\prime \Psi_0 -3\overline{\mathcal{Q}}\Psi_2,  \label{I1}
\end{equation}
\begin{equation}
aD\Psi_2 +b\Delta\Psi_2 -c \eth\Psi_2 -\cbar\eth \Psi_2= -2\overline{\mathcal{Q}}^\prime \Psi_1 - 2\overline{\mathcal{Q}}\Psi_3,  \label{I2}
\end{equation}
where
\begin{equation}
\mathcal{Q}=\ethbar b +\taub b,
\end{equation}
\begin{equation}
\mathcal{Q}^\prime=\eth a - \pib a - \nub b,
\end{equation}
\begin{equation}
\mathcal{P}= -Da + \pib c + \pi \cbar,
\end{equation}
\begin{equation}
2i\mathcal{S}= \ethbar \cbar - \eth c.
\end{equation}

The equations (\ref{K1})-(\ref{I2}) will be expanded in polyhomogeneous series, and then solved order by order. This analysis will yield Killing vector fields that are consistent with the polyhomogeneous asymptotically flat spacetimes of \cite{xiv}, \cite{javk98a}, \cite{javk98c}. And more importantly, it will also yield constraints for the quantities that are initial data at $\N_0$, and $\Z$ (i.e. $\Psi_0$, and $\sigma_{2,0}$, $\Psi_1^{4,0}$, $\Psi_2^{3,0}$).

It will be  assumed that the coefficients in the Killing field can be expanded as polyhomogeneous series.
\begin{eqnarray}
a &=&a_{-1}r+a_0+a_1r^{-1}+a_2r^{-2}+a_3r^{-3}+ \ldots, \\
b &=&b_{-1}r+b_0+b_1r^{-1}+b_2r^{-2}+b_3r^{-3}+ \ldots, \\
c &=&c_{-1}r+c_0+c_1r^{-1}+c_2r^{-2}+c_3r^{-3}+ \ldots,
\end{eqnarray}
where the $a_i$, $b_i$, and $c_i$ are polynomials in $z=\ln r$.

\section{The asymptotic Killing vector.}

Solving the Killing equations to the leading order in $1/r$ yields the asymptotic form of the Killing vector fields that are compatible with the asymptotically flat spacetimes under consideration. The expansions yield

\begin{eqnarray}
\partial _rb &=&0, \label{A1}\\
\partial _ua_{-1} &=&0, \label{A2} \\
\ethbar_1 c_{-1} &=&0, \label{A3} \\
\partial _ub+a_{-1}+a_{-1}^{\prime} &=&0, \label{A4}\\
\eth_1 c_{-1}+\ethbar_1 \cbar_{-1} &=&2a_{-1}, \label{A5}\\
c_{-1}^{\prime } &=&0, \label{A6}\\
\partial _uc_{-1} &=&0. \label{A7}
\end{eqnarray}
From equations (\ref{A1}) and (\ref{A6}) together with (\ref{A4}) one sees that

\begin{equation}
\#a_{-1}=\#c_{-1}=0,
\end{equation}
and $b$ is independent of $r$. Hence, there are no logarithmic terms at this order. From $\ethbar_1 c_{-1}=0$ (using lemma 1) one finds that $c_{-1}$ has to be a linear combination of $l=1$ spherical harmonics:
\begin{eqnarray}
c_{-1} &=&\sum^1_{m=-1}(-1)^{m+1}A_m\left( _{-1}Y_{1,-m}\right) , \\
\cbar_{-1} &=&\sum^1_{m=-1}\overline{A}_m\left( _1Y_{1,m}\right),
\end{eqnarray}
where $A_m \epsilon$ \Complex. Using the properties of spin weighted spherical harmonics one can readily find that

\begin{equation}
a_{-1}=\frac 12\sum^1_{m=-1}\left\{\overline{ A}_m+(-1)^m{A}_{-m}\right\} (_0Y_{1,m}),
\end{equation}
and hence

\begin{equation}
b=-\frac u2\sum^1_{m=-1}\left\{ \overline{A}_m+(-1)^m{A}_{-m}\right\}
(_0Y_{1,m})+\alpha (\theta ,\varphi ),
\end{equation}
where $\alpha(\theta,\varphi)$ is an arbitrary integration function. The asymptotic form of the Killing vector field is therefore

\begin{equation}
\xi ^\mu =\left( -a_{-1}u+\alpha (\theta ,\varphi ),a_{-1}r,-\frac 1{\sqrt{2}%
}(\cbar_{-1}+c_{-1}),-\frac{i\csc \theta }{\sqrt{2}}(\cbar_{-1}-c_{-1})\right).
\end{equation}
Hence, all Killing vector fields compatible with asymptotic flatness can be
constructed by providing the 3 complex numbers $A_{-1}$, $A_0$, $A_1$ and
the function $\alpha (\theta ,\varphi )$. Polyhomogeneity adds nothing new at this order.
In the axisymmetric case, it has been shown that the integration function $\alpha(\theta,\varphi)$ can be removed with a supertranslation. This suggests that a similar thing can be done in the general case. If one lets $u^{\prime }=u+\beta $, then the Killing vector field will transform as

\begin{eqnarray}
\xi ^{\prime u} &=&\xi ^u-c_{-1}\eth_1 \beta -\cbar_{-1}\ethbar_1   
\beta , \\
\xi ^{\prime r} &=&\xi ^r, \\
\xi ^{\prime \theta } &=&\xi ^\theta , \\
\xi ^{\prime \varphi } &=&\xi ^\varphi.
\end{eqnarray}
In order to remove $\alpha$, one needs to find a solution $\beta$ to the partial differential equation
\begin{equation}
\alpha =c_{-1}\eth_1 \beta +\cbar_{-1}\ethbar_1 \beta. \label{supert}
\end{equation}
As is shown in the appendix A, it is always possible to construct a $\beta$ that satisfies equation (\ref{supert}). This shows that the function $\alpha (\theta ,\varphi )$ is associated with supertranslational Killing vector fields, i.e. fields of the form

\begin{equation}
\xi _{\sup }^\mu =\left( \alpha(\theta,\varphi) ,0,0,0\right).
\end{equation}

\subsection{Axial symmetry}

In an axially symmetric spacetime, one expects all the quantities to be independent of the coordinate $\varphi$. This fact constrains the expansions in spherical harmonics  of functions over $S^2$, as the only harmonics that are independent of $\varphi$ are those with $m=0$. Hence, in an axisymmetric spacetime  a quantity of spin weight $s$, $\eta$ will be of the form
\begin{equation}
\eta = \sum_{k=s} \eta_{k} (_s Y_{k,0}).
\end{equation}
The Killing vector for axial symmetry is

\begin{equation}
\eta ^\mu =(0,0,0,1),
\end{equation}
so that 
\begin{equation}
c=\frac{-i\sin \theta }{\sqrt{2}}r,
\end{equation}
and 
\begin{equation}
a=b=0.
\end{equation}

Now, we note that the vector $n^\mu$ can be chosen so that at the origin, its spacelike projection lies in a plane containing the axis of symmetry. Due to the axial symmetry of the spacetime, the projection will remain in this plane. Therefore $n^\mu$ will have no components in the direction of $\partial_\varphi$. Hence $C^3=0$.
The null vectors $m^\mu$ and $\mbar^\mu$ were constructed so that they span $\mbox{T}(S_{u,r})$. The dimension of $\mbox{T}(S_{u,r})$ is 2, but the vector $m^\mu$ depends on 4 real functions of $(u,r,\theta)$. The 2 extra functions are related to the freedom of performing a spin $m^\mu \mapsto e^{i\vartheta}m^\mu$. From $m^\mu$ and $\mbar^\mu$ one can construct two real vectors
\begin{equation}
e^\mu_2=\frac{1}{\sqrt{2}}\left( m^\mu+\mbar^\mu \right)=(0,0,\Re \, \xi^2, \Re \, \xi^3),
\end{equation}
\begin{equation}
e^\mu_3=\frac{i}{\sqrt{2}}\left( m^\mu-\mbar^\mu\right)= (0,0,\Im \, \xi^2,\Im \, \xi^3).
\end{equation}
The using the freedom left in the spin it is possible to set $\Re \, \xi^3=\Im \, \xi^2=0$. In this way $e^\mu_3$ will lie in the direction of $\partial_\varphi$, and  $\xi^2$ will be real and $\xi^3$ will be pure imaginary. After this choice, all the tetrad freedom will have been removed and all the spin weighted quantities will have lost their spin weight. If $\xi^2$ is real, and $\xi^3$ is pure imaginary then from the frame equation (Fb)
\begin{equation}
D\xi^\alpha=\rho\xi^\alpha + \sigma \xib^\alpha,
\end{equation}
one deduces that $\sigma$ has to be real, and hence from the NP field equation (b)
\begin{equation}
D\sigma=2\rho\sigma +\Psi_0,
\end{equation}
$\Psi_0$ is real. These results are summarized in the following lemma.

\begin{lemma}
For an axially symmetric spacetime, it is possible to choose the tetrad so that $C^3=0$, $\xi^2$, $\sigma$, $\Psi_0$ are real, and $\xi^3$ is pure imaginary.Under this choice, there will be no spin freedom left.
\end{lemma} 

\subsection{Boost-rotation symmetric spacetimes}

If the spacetime is axisymmetric ($\eta^\mu$), and happens to possess another Killing vector ($\xi^\mu$), then the two Killing vector fields will form an Abelian algebra \cite{bicak} .

If one assumes that the spacetime is axisymmetric, then the coefficients $a_{-1}$, $b$, $c_{-1}$ and $\cbar_{-1}$ for the other Killing vector, $\xi^\mu$ are constrained to be
\begin{equation}
c_{-1}= A_0 \sin \theta,
\end{equation}
\begin{equation}
\cbar_{-1}= \overline{A}_0\sin\theta,
\end{equation}
\begin{equation}
a_{-1}= \frac{1}{\sqrt{2}} (A_0 + \overline{A}_0) \cos\theta,`
\end{equation}
\begin{equation}
b=-\frac{u}{\sqrt{2}} (A_0 + \overline{A}_0) \cos\theta.
\end{equation}
Therefore, the asymptotic form of the extra Killing vector field has to be
\begin{equation}
\xi ^\mu =(-\Re(A_0)u\cos \theta +\alpha (\theta ),\Re(A_0)r\cos \theta ,-\Re(A)\sin \theta ,-\Im(A_0)).
\end{equation}
With any loss of generality one can set $\Im(A_0)$ equal to zero as the vector $(0,0,0,\Im(A_0))$ is just a multiple of the axial Killing field. Hence, Bi\v{c}\'{a}k \& Schmidt's \cite{bicak} result has been recovered. The resulting vector is known as the boost-rotation Killing vector.

\section{Constraints due to the presence of a Killing field}

\subsection{General remarks}

As has been seen in the previous section, the expansion of the Killing equations to order $-1$ in $1/r$ yields the asymptotic form of the Killing vector field. If one carries the expansions to further orders, one will expect to observe the interaction of the symmetry of the spacetime with the quantities that are initial data at a given null hypersurface $\N_0$ through the integrability conditions of the Killing equations. The presence of the Killing vector field will require the initial data to satisfy some constraint equations that will be derived from the integrability conditions. The constraint equations for $\sigma_2$, $\Psi_0$, $\Psi_1^4$ and $\Psi_2^3$ can be deduced from the integrability conditions ((IS), (I0), (I1), (I2)), equations (\ref{IS}), (\ref{I0}), (\ref{I1}), (\ref{I2}). Looking at the integrability conditions it is not difficult to see that the generic form of the constraint equations will be:

\begin{equation}
b\partial_u X - c_{-1} \eth_1 X -\cbar_{-1} \ethbar_1 X + H X =Q, \label{generic}
\end{equation}
where $H$ and $F$ depend on quantities that have been calculated at previous orders in the expansions. A number of observations can be made from this equation. The constraint equation (\ref{generic}) is a linear partial differential equation for the (in principle) complex quantity $X$. The domain of the solutions to this equation is at least a piece of $\scri=$\Real$\times S^2$. This fact suggests that one could use separation of variables in order to try to solve the equation when the non-homogeneous term $Q$ is not present. Hence, let us assume that

\begin{equation}
X(u,\theta,\varphi)=U(u)\Omega(\theta,\varphi).
\end{equation}
Substitution into the homogeneous part of equation (\ref{generic}) yields 
\begin{equation}
(-ua_{-1})U^{\prime}\Omega-c_{-1}U\eth_1\Omega -\cbar_{-1}U \ethbar_1\Omega + H U\Omega=0.
\end{equation}
Dividing by $U\Omega$, and collecting terms one gets
\begin{equation}
-u\frac{U^{\prime}}{U}=\frac{1}{a_{-1}} \left(  \frac{c_{-1}\eth_1\Omega+\cbar_{-1} \ethbar_1\Omega}{\Omega} -H \right).
\end{equation}
Using the  classical argument of separation of variables, one sees that the left-hand side depends only on $u$, while the right-hand side depends only on the angular coordinates $(\theta,\phi)$. Hence each side has to be equal to a constant $\Lambda=\Lambda_1+i\Lambda_2$ ($\Lambda_1$, $\Lambda_2\epsilon$\Real). Therefore
\begin{equation}
U^{\prime}+\Lambda uU=0, \label{udependence}
\end{equation}
and
\begin{equation}
c_{-1}\eth_1\Omega +\cbar_{-1}\ethbar_1\Omega-(H+a_{-1}\Lambda)\Omega=0 \label{angdependence}
\end{equation}
Equation (\ref{udependence}) can be solved readily yielding
\begin{equation}
U(u)=C_{\Lambda_1,\Lambda_2} e^{-\frac{1}{2}(\Lambda_1+i\Lambda_2)u^2},
\end{equation}
where $C_{\Lambda_1,\Lambda_2}$ is an integration constant. The structure of the function $U$ is extremely suggestive. The number $\Lambda_1$ is clearly a damping parameter, and $\Lambda_2$ is a frequency (which we will assume to be positive). In principle there is no mathematical restriction to the values $\Lambda$ can take, but on physical grounds one would like to restrict them to the first quadrant of the complex plane ($\Lambda_1$, $\Lambda_2 \geq0$) as one expects the ``background'' gravitational field to damp the propagation of gravitational radiation.

The solution of equation (\ref{angdependence}) will be (in principle) an infinite series of spin-weight $s$ spherical harmonics. A general solution for the constraint equation (\ref{constraintnews}) is given by integrating over the values of the parameter $\Lambda$:

\begin{equation}
X=\int^{\infty}_0 \int^{\infty}_{0} C_{\Lambda_1,\Lambda_2}e^{-\frac{1}{2}(\Lambda_1+i\Lambda_2)u^2}\Omega_{\Lambda_1,\Lambda_2}(\theta,\varphi) \d\Lambda_2 \d\Lambda_1.
\end{equation}

Another interesting observation on the behaviour of solutions to equation (\ref{generic}) can be obtained from writing the $\eth_1$ and $\ethbar_1$ operators as ordinary partial derivatives. If $X$ is a quantity of spin-weight $s$ then
\begin{eqnarray}
\eth_1 X= \frac{\sin^s\theta}{\sqrt{2}} \left( \partial_\theta -i\csc\theta \partial_\varphi \right) ( \sin^{-s}\theta X), \\
\ethbar_1 X =\frac{\sin^{-s}\theta}{\sqrt{2}} \left( \partial_\theta +i\csc\theta \partial_\varphi \right) ( \sin^{s}\theta X),
\end{eqnarray}
and the differential equation (\ref{generic}) takes the form

\begin{equation}
(-a_{-1}u)\partial_u X -\frac{1}{\sqrt{2}} (\cbar_{-1}+c_{-1})\partial_\theta X -\frac{i\csc\theta}{\sqrt{2}}(\cbar_{-1}-c_{-1}) \partial_\varphi X +\left(H-\frac{s\cot\theta}{\sqrt{2}}(\cbar_{-1}-c_{-1})\right) X  = Q. \label{standard}
\end{equation}
Therefore the tangent vector field to the characteristics of equation (\ref{standard}) is given by

\begin{equation}
\chi^\mu= \left( -a_{-1}u+\alpha, -\frac{1}{\sqrt{2}}(\cbar_{-1}+c_{-1}), -\frac{i\csc \theta}{\sqrt{2}}(\cbar_{-1}-c_{-1}) \right),
\end{equation}
that is, the restriction of the Killing vector field $\xi^\mu$ to $\scri$, $\xi^\mu|_{\scri}$. Hence, the characteristics of equation (\ref{generic}) are the orbits of $\xi^\mu|_\scri$. The structure of the equations (\ref{generic}) and (\ref{standard}) is very suggestive as it resembles that of the continuity equations of mechanics of continuous media: $\rho_t+ \mbox{div}J=F$, where $F$ describes a source or a sink of the quantity $\rho$, and $J$ its flux.

From the theory of first order linear partial differential equations, one knows that equation (\ref{standard}) is equivalent to a system of 4 (non-linear) ordinary differential equations:
\begin{eqnarray}
\frac{\d u}{\d t}&=&-a_{-1}u, \\ \label{ode1}
\frac{\d \theta}{\d t}&=& -\frac{1}{\sqrt{2}}(\cbar_{-1}+c_{-1}), \\ \label{ode2}
\frac{\d \varphi}{\d t}&=& -\frac{i\csc \theta}{\sqrt{2}}(\cbar_{-1}-c_{-1}), \\ \label{ode3}
\frac{\d X}{\d t}&=& -\left(H-\frac{s\cot\theta}{\sqrt{2}}(\cbar_{-1}-c_{-1})\right) X + Q, \label{ode4}
\end{eqnarray}
with initial data $u(0)=u_0$, $\theta=\xi$, $\varphi=\eta$ ($\xi\,\epsilon\,[0,\pi]$, $\eta \, \epsilon \,[0,2\pi)$), $X(0)=f(\xi,\eta)$. The first 3 equations yield the characteristics/orbits of the restriction of the Killing vector field to $\scri$. The function $f$ is therefore by construction an invariant along the orbits. Note that because we are solving the Killing equations and their integrability conditions order by order, then the constraint equations will form a hierarchy. It will be necessary to solve all the constraints up to order say $k$ in $1/r$ in order to be able to solve the constraints to order $k+1$. This procedure will give rise to a set of functionally-independent invariants along the orbits from which it will be possible to construct all the spin-weighted quantities.

The characteristic curves are represented by the mapping $(t,\xi,\eta) \mapsto (u,\theta,\varphi)$ ($t$ is the parameter of the curve, and $\xi$ and $\eta$ are the coordinates of the end point of the curve at the initial cut of $\scri$) which can be inverted as long as its Jacobian is different from zero. If this is indeed the case, one can solve equation (\ref{ode4}) to get the solution of the original partial differential equation. The solution to the homogeneous part of equation(\ref{ode4}) can be formally written as:

\begin{equation}
X_h= f(\xi,\eta) e^{-\int_{\mathcal{C}} \left(H-\frac{1}{\sqrt{2}}s\cot\theta(\cbar_{-1}-c_{-1})\right)}, \label{xh}
\end{equation}
where $\xi$ and $\eta$ are functions of $(u,\theta,\varphi)$ as discussed before, and the integration has to be understood as a line integral along the (unique) characteristic that goes through $(u,\theta,\phi)$ and which can be retrodicted up to the initial cut ($\Z_0=\scri\cap\N_0$). A particular solution to the  equation (\ref{ode4}) is given by

\begin{equation}
X_p=e^{ -\int_{\mathcal{C}} \left(H-\frac{1}{\sqrt{2}}s\cot\theta(\cbar_{-1}-c_{-1})\right) } \int_{\mathcal{C}} Q e^{ \int_{\mathcal{C}} \left(H-\frac{1}{\sqrt{2}}s\cot\theta(\cbar_{-1}-c_{-1})\right) }. \label{xp}
\end{equation}
The complete solution will be therefore
\begin{equation}
X=X_h + X_p,
\end{equation}
where the homogeneous term ($X_h$) is associated with the propagation of the initial data $f(\theta,\phi)$, which will be damped due to the ``interaction'' of the quantity $X$ with the background (represented by the term $(H-\frac{1}{\sqrt{2}}s\cot\theta(\cbar_{-1}-c_{-1}))X$). The non-homogeneous part will account for the added effects (in time) of the source/sink.

\subsection{Constraint on the news function}

The shear ($\sigma$) is a quantity of great physical interest as the derivative of its leading term with respect to the retarded time $u$ ($\dot{\sigma}_{2,0}$), known as the \emph{news function}, determines whether or not the spacetime is radiative. The outgoing radiation field for the asymptotically flat spacetime is determined by the leading term of $\Psi_4$, which itself depends on the news function ($\Psi_4^{1,0}=-\ddot{\sigmab}$).
In order to obtain a constraint equation for the leading term of the shear ($\sigma$) one has to expand the integrability condition (IS) (equation (\ref{IS})) to order 2 in $1/r$. One directly finds that:

\begin{equation}
\dot{\sigma}_{2,0}b- c_{-1} \eth_1\sigma_{2,0}- \cbar_{-1} \ethbar_1\sigma_{2,0}-(a_{-1}+\eth_1c_{-1}-\ethbar_1\cbar_{-1})\sigma_{2,0} =0 \label{constraintsigma}.
\end{equation}
Differentiating the latter with respect to $u$, one obtains a constraint equation for the news function ($\dot{\sigma}_{2,0}$),

\begin{equation}
\ddot{\sigma }_{2,0}b-c_{-1}\eth_1 \dot{\sigma }_{2,0}-\cbar_{-1}\ethbar_1 \dot{\sigma }_{2,0}-2(\eth_1 c_{-1})\dot{\sigma }_{2,0}=0.  \label{constraintnews}
\end{equation}
Once these constraint equations have been solved, one can proceed to solve the Killing equations (\ref{K1})-(\ref{K7}) at order 0 in $1/r$,

\begin{eqnarray}
\dot{a}_0-\frac 12\left( a_{-1}^{\prime }+a_{-1}\right) &=&0,
\label{B1} \\
\ethbar_1  c_0+\ethbar_2 c_{-1} &=&-\sigmab_2a_{-1}+\lambda_1b,  \label{B2} \\
a_0^{\prime } &=&0,  \label{B3} \\
\eth_1 c_0+\ethbar_1\cbar_0+\ethbar_2\cbar_{-1}+\eth_2 c_{-1} &=&2a_0-b,  \label{B4} \\
\ethbar_1 b &=&c_0-c_0^\prime-\sigmab_2\cbar_{-1},
\label{B5} \\
\dot{\cbar}_0+\eth_1 a_{-1} &=&\lambdab_1 c_{-1},  \label{B6}
\end{eqnarray}
where $\lambda_1=\dot{ \sigmab}_{2,0}$, and $\eth_2=-\sigma_{2,0}\ethbar_1$.  Note that because of $\Psi_0=\O(r^{-4} \ln^{N_4}r)$, then $\#\sigma_2=0$, and hence $\sigma_2=\sigma_{2,0}$.  The coefficients $c_0$ and $a_0$ can be calculated from equations (\ref{B5}) and (\ref{B4}) respectively. It can be easily seen that $\#a_0=\#c_0=0$. 

\subsubsection{Supertranslational Killing field.}

It is known that an axisymmetric spacetime that admits a supertranslational Killing vector is necessarily non-radiative \cite{ashtekar1}, \cite{ashtekar2}, \cite{bicak}. Using equation (\ref{constraintsigma}) it is easy to show that this assertion is still valid even if we remove the hypothesis of axial symmetry.  For a supertranslational Killing vector field
we have that
\begin{eqnarray}
c_{-1}&=&0, \\
a_{-1}&=&0, \\
b&=&\alpha(\theta,\varphi).
\end{eqnarray}
Therefore the constraint equation for $\sigma_{2,0}$ is

\begin{equation}
\dot{\sigma}_{2,0}b=0.
\end{equation}

This clearly shows that if $\alpha \neq 0$ then necessarily $\dot{\sigma}_{2,0}=0$. So we obtain the following proposition.

\begin{proposition}
If an asymptotically flat spacetime admits a supertranslational Killing
vector field then it is non-radiative ($\dot{\sigma}_{2,0}=0$ for all retarded times).
\end{proposition}

\subsubsection{Boost-rotation symmetry.}

From the discussion in section (2.7), the leading terms of the boost-rotation Killing vector are given by
\begin{eqnarray}
c_{-1}&=& A_0 \sin \theta, \\
a_{-1}&=&\sqrt{2} A_0 \cos\theta, \\
\eth_1c_{-1}&=& \sqrt{2}A_0 \cos \theta, \\
b&=& -u\sqrt{2} A_0 \cos \theta, 
\end{eqnarray}
where $A_0\epsilon$\Real. If one fixes the tetrad in the way prescribed by lemma 2 in section (2.6), then $\sigma_{2,0}$ is real, and the constraint equations(\ref{constraintsigma}) and (\ref{constraintnews})  simplify to

\begin{equation}
(u\cot \theta )\partial_u(\sigma_{2,0})+\partial_{\theta}(\sigma_{2,0})+( \cot\theta)\sigma_{2,0}=0, \label{boostsigma}
\end{equation}

\begin{equation}
(u\cot \theta )\partial_u(\dot{\sigma}_{2,0})+\partial_{\theta}(\dot{\sigma}_{2,0})+(2 \cot\theta)\dot{\sigma}_{2,0}=0. \label{boostnews}
\end{equation}
Both partial differential equations can be solved easily using the method of characteristics. The last equation is a particular case of the equation $u\cot \theta \partial_ux+\partial _\theta x+k\cot \theta x=H(u,\theta )$. This kind of equation will appear many times more, therefore, a study of its solutions is made in appendix B. The general solution of (\ref{boostsigma}) is

\begin{equation}
\sigma_{2,0}=(\csc\theta)G(\frac{\sin\theta}{u}),
\end{equation}
and that of equation(\ref{boostnews}) is

\begin{equation}
\dot{\sigma }_{2,0}=\frac 1{u^2}F_1\left( \frac{\sin \theta }u\right),
\end{equation}
where $G$ and $F_1$ are arbitrary functions of $\sin \theta /u$, and $F_1=-G^\prime$. Hence, we have recovered the results of \cite{bicak}.

\subsection{Constraints on $\mbox{Re} \, \Psi_2^{3,0}$ and $\Psi _0^{4,N_4}$}

The next step in our study is to obtain constraint equations for $\mbox{Re} \, \Psi_2^{3,0}$ (the mass aspect of the spacetime), and $\Psi_0^4$ (the coefficient that give rise to the logarithmic terms in the expansions). The constraint equation for $\mbox{Re} \, \Psi_2^{3,0}$ can be deduced either from the constraint equation (\ref{I2}) or from the expansions of the Killing equations at order 1 in $1/r$:

\begin{eqnarray}
&-\nu_2\cbar_{-1}-\nub_2c_{-1}+\dot{a}
_1+Q_1a_{-1}+C_2^\alpha \partial _\alpha a_{-1}+(\gamma _2+\gammab 
_2)a_{-1}=0,&  \label{C1} \\
&\sigmab_3 a_{-1}+\sigmab_2a_0+\ethbar_1 c_1+\ethbar_2c_0
+\ethbar  _3c_{-1}=\lambda _2b,&  \label{C2} \\
&C_2^\alpha \partial _\alpha b-(\gamma _2+\gammab _2)b=a_1,&  \label{C3} \\
&\ethbar_1\cbar_1+\ethbar_2\cbar_0+\ethbar_3\cbar_{-1}
+\eth_1 c_1+\eth_2c_0+\eth_3 c_{-1}-2\mu _2b+2\rho _3a_{-1}=2a_1,&  \label{C4}\\
&\sigmab _3\cbar_{-1}+\sigmab  _2\cbar
_0-2c_1+c_1^{\prime }+\rho _3c_{-1}+\ethbar _2b-\pi _2b=0,&
\label{C5} \\
&Q_1\cbar_{-1}-\frac 12\cbar_0^{\prime }+\tau _2a_{-1}+\eth
_2a_{-1}+\eth_1 a_0-\mu _2\cbar_{-1}+\frac 12\cbar_0+
\dot{\cbar}_1& \nonumber \\
&+C_2^\alpha \partial _a\cbar_{-1}-(\gamma _2-
\gammab _2)\cbar_{-1}-\lambda_2c_{-1}-\lambdab
_1c_0=0,&  \label{C6} 
\end{eqnarray}
where \cite{stewart}, \cite{javk98c}

\begin{eqnarray}
&\sigma=\sigma_{2,0}r^{-2}+\sigma_3[N_4]r^{-3}+\O(r^{-4}\ln^{N_5}r),& \\
&\nu =\ethbar_1  Mr^{-2}+\O(r^{-3}\ln^{N_4+1}r),&\\
&Q=-\frac 12-Mr^{-1}+\O(r^{-2}\ln^{N_4+1}r),&\\
&C_2^\alpha \partial _\alpha =-\left[ (\eth_1 \sigmab_{2,0})\delta
_1+(\ethbar_1 \sigma _{2,0})\overline{\delta }_1\right] r^{-2}+\O(r^{-3}\ln^{N_4+1}r),&\\
&\gamma _2+\gammab _2=-Mr^{-2}+\O(r^{-3}\ln^{N_4+1}r),&
\end{eqnarray}
and  $M=\mbox{Re}\Psi_2^{3,0}=\frac 12(\Psi _2^{3,0}+\overline{\Psi }_2^{3,0})$. From (\ref{C3}) one sees that $\#a_1=0$, and from equation (\ref{C5}) $\#c_1=\#\sigma_3=N_4$. Using all these results equation (\ref{C1}) reads

\begin{equation}
-2Ma_{-1}-\cbar_{-1}\ethbar_1 M+\dot{a}_1-\eth_1 \sigmab
_{2,0}\eth_1 a_{-1}-\ethbar_1\sigma _{2,0}\ethbar_1 a_{-1}-c_{-1}\eth_1
M=0.
\end{equation}

From equation (\ref{C3}) one obtains
\begin{equation}
a_1=Mb-\eth_1 \sigmab_{2,0}\eth_1 b-\ethbar_1 \sigma _{2,0}\ethbar_1 b.
\end{equation}

Combining the two equations, and recalling that $\dot b=-a_{-1}$ one finally obtains the desired constraint equation,

\begin{equation}
b\dot{M}-\cbar_{-1}\ethbar_1M-c_{-1}\eth_1M+3\dot{b}M=
(\eth_1 \dot{\sigmab}_{2,0}\eth_1 b+\ethbar_1
\dot{\sigma }_{2,0}\ethbar_1 b).  \label{massconstraint}
\end{equation}
Note that this constraint equation has a sink term ($(\eth_1 \dot{\sigmab}_{2,0}\eth_1 b+\ethbar_1\dot{\sigma }_{2,0}\ethbar_1 b)$) that depends on the news function ($\dot{\sigma}_{2,0}$), as one may expect from the Bondi mass formula (see e.g. \cite{stewart}). Once $M$ is determined, the coefficient $a_1$ is readily found. Using the integrability condition (I0) (equation(\ref{I0})) at order 4 in $1/r$ one can easily deduce the constraint equation for $\Psi_0^4$,

\begin{equation}
c_{-1}\eth_1\Psi_0^4 + \cbar_{-1}\ethbar_1\Psi_0^4+2\eth_1c_{-1}\Psi_0^4=-a_{-1}\Psi_0^{4 \prime};
\end{equation}
in particular the coefficient of the highest $\ln r$ power ($N_4$) should satisfy:
\begin{equation}
c_{-1}\eth_1 \Psi _0^{4,N_4}+\cbar_{-1}\ethbar_1\Psi_0^{4,N_4}+2\eth c_{-1}\Psi _0^{4,N_4}=0. \label{constraintpsi04}
\end{equation}
Note that there are no derivatives with respect to the retarded time, as $\Psi_0^4$ is a constant of motion \cite{javk98a}.

\subsubsection{Boost rotation symmetry.}

It can be shown that in the case of boost-rotation symmetric spacetimes, the constraint equation for the mass aspect $M$ reduces to the one found by Bi\v{c}\'{a}k \& Pravdov\'{a} \cite{bicak3}:

\begin{equation}
u\cot\theta \partial_u M +\partial_\theta M + 3\cot\theta M= u \partial_u (\partial_\theta \sigma_{2,0} + 2 \cot\theta \sigma_{2,0}).
\end{equation}
From appendix B we learn that the homogeneous part of the solution of the previous equation will be of the form $M_h=u^{-3}F(u^{-1}\sin\theta)$. A discussion of the behaviour of the solutions to the constraint equation, and their relation with the Bondi mass of the spacetime can be found in \cite{bicak3}.

The constraint equation for $\Psi_0^{4,0}$ (in a ``minimal'' polyhomogeneous spacetime) reduces to an ordinary differential equation ($\Psi_0^4$ is a constant of motion \cite{javk98a}),

\begin{equation}
\frac{\d}{\d \theta}\Psi^{4,0}_0 +2 \cot\theta\Psi^{4,0}_0=0 \label{boostpsi04},
\end{equation}
whose solution is
\begin{equation}
\Psi _0^{4,0}=C_1\csc ^2\theta.
\end{equation}
Therefore $\Psi_0^4$ is singular at $\theta=0$, $\pi$. Note that $\Psi_0$ is an invariant of the boost-rotation symmetric spacetime, as all the freedom in the tetrad has been removed. Hence, there are singularities at the ``north'' and ``south'' poles. From this analysis one concludes that a polyhomogeneous boost-rotation symmetric spacetime with $\Psi_0=\O(r^{-4}\ln^{N_4}r)$ can only have a ``local'' $\scri$, i.e. $\scri$ is not isomorphic to $S^2\times$\Real \ \cite{ashtekar1}, \cite{ashtekar2}. If one wishes to have a boost-rotation symmetric spacetime with at least a ``piece'' of $\scri$ ($\scri\simeq S^2\times$\Real) then one must set $\Psi_0^4=0$. Note as well that $\ethbar_1 \Psi_0^{4,0}=C_1\ethbar_1 (\csc^2\theta)=0$. This fact will simplify future calculations.

\subsection{Constraint on $\Psi_1^{4,0}$ and $\Psi _0^{5}$}

The integrability condition (I1) (equation(\ref{I1})) expanded at order 4 in $1/r$ yields the constraint equation for $\Psi_1^{4,0}$, while the constraint equation for $\Psi _0^{5}$ can be deduced from the integrability condition (I0) (equation (\ref{I0})) at order 5 in $1/r$.  In the theory of multipole expansions of stationary spacetimes these coefficients are closely related to the dipole moment and the quadrupole moment respectively \cite{hansen}, \cite{kundu}. For radiative spacetimes, these relations do not hold anymore; however $\Psi_1^{4,0}$ is an indispensable ingredient of all the definitions of angular momentum for asymptotically flat spacetimes \cite{dray}, \cite{penrose}.

The constraint equations are respectively

\begin{equation}
b\dot{\Psi}_1^4-c_{-1}\eth_1\Psi_1^4 -\cbar_{-1}\ethbar_1 \Psi_1^4 -(2a_{-1}+\eth_1c_{-1})\Psi_1^4= -a_{-1}\Psi_1^{4 \prime} -3\eth_1b \Psi_2^{3}, \label{Psi14}
\end{equation}
and 
\begin{eqnarray}
&b\dot{\Psi}_0^5 - c_{-1}\eth_1 \Psi_0^5 -\cbar_{-1}\ethbar_1\Psi_0^5-(a_{-1}+2\eth_1c_{-1})\Psi_0^5=& \nonumber \\
&-a_{-1}\Psi_0^{5 \prime}+(\frac{1}{2}b-a_0)\Psi_0^{4\prime}+(4a_0-2b+2\tau_2c_{-1}+2\taub_2\cbar_{-1} +\eth_1c_0 +\eth_2c_{-1}-\ethbar_1\cbar_0 -\ethbar_2\cbar_{-1})\Psi_0^4& \nonumber \\
& + c_0\eth_1\Psi_0^4 +\cbar_0\ethbar_1\Psi_0^4 +c_{-1}\eth_2\Psi_0^4 + \cbar_{-1}\ethbar_2\Psi_0^4 -4\eth_1b\Psi_1^4.&
\end{eqnarray}
Now, from \cite{javk98a} one knows that $\Psi_1^{4 \prime}=\ethbar_1 \Psi_0^4$ so that $(N_4+1)\Psi_1^{4,N_a+1}=\ethbar_1\Psi_0^{4,N_4}$; therefore the only new constraint equation one can deduce from (\ref{Psi14}) is that for $\Psi_1^{4,0}$ (the equations for the other coefficients are satisfied identically), therefore:

\begin{equation}
b\dot{\Psi}_1^{4,0}-c_{-1}\eth_1\Psi_1^{4,0} -\cbar_{-1}\ethbar_1 \Psi_1^{4,0}  -(2a_{-1}+\eth_1c_{-1})\Psi_1^{4,0}= -a_{-1}\ethbar_1\Psi_0^{4,0} -3\eth_1b \Psi_2^{3,0}. \label{psi140}
\end{equation}
This constraint is valid both for polyhomogeneous and non-polyhomogeneous spacetimes. It can be regarded as describing a process of transformation of dipole moment (angular momentum) into mass monopole momment due to the gravitational radiative process.

 For a non-polyhomogeneous spacetime the constraint for the coefficient $\Psi_0^{5,0}$ reduces to,

\begin{equation}
b\dot{\Psi}_0^{5,0} - c_{-1}\eth_1 \Psi_0^{5,0} -\cbar_{-1}\ethbar_1\Psi_0^{5,0}-(a_{-1}+2\eth_1c_{-1})\Psi_0^{5,0}=-4\eth_1b\Psi_1^{4,0},
\end{equation}
while for a ``minimal'' polyhomogeneous spacetime the leading coefficient ($\Psi_0^{5,0}$) should satisfy
\begin{equation}
b\dot{\Psi}_0^{5,1} - c_{-1}\eth_1 \Psi_0^{5,1} -\cbar_{-1}\ethbar_1\Psi_0^{5,1}-(a_{-1}+2\eth_1c_{-1})\Psi_0^{5,1}=-4\eth_1b\ethbar_1\Psi_0^{4,0}.
\end{equation}
In a similar way to what happened with the constraint equation for $\Psi_1^{4,0}$, these two last equations can be interpreted as describing a process of interchange of mass quadrupole moment into dipole moment (angular momentum). Note that for the ``minimal'' polyhomogeneous spacetime there will be  two coefficients associated with the quadrupole, the logarithmic one being dominant far away from the source.

As in the previous subsections, once the constraint equations have been solved, one can easily find the coefficients $a_2$ and $c_2$ from the Killing equations (K5) and (K6) (equations (\ref{K5}) and (\ref{K6})) at order 2 in $1/r$:
\begin{eqnarray}
&\ethbar_3\cbar_0+\ethbar_2\cbar_1+\ethbar_1\cbar_2+\ethbar_4\cbar_{-1}+2\rho _3a_0+2\rho _4a_{-1}-2\mu _3b&  \nonumber \\
&+\eth _4c_{-1}+\eth _3c_0+\eth _2c_1+\eth_1c_2 =2a_2,& \label{D4}
\end{eqnarray}
\begin{equation}
\rho _3c_0+\rho _4c_{-1}-3c_2+\ethbar_3b+c_2^{\prime }-\pi _3b+
\sigmab _{2,0}\cbar_1+\sigmab  _4\cbar_{-1}+
\sigmab _3\cbar_0=0.  \label{D5}
\end{equation}
From these equations one can see that $\#a_2=\#\gamma_3=N_4+1$ and $\#c_2=N_5$. 

\subsubsection{Boost rotation symmetry.}

The constraint equations for $\Psi_1^{4,0}$ are the same for non-polyhomogeneous and ``minimal'' polyhomogeneus boost-rotation symmetric spacetimes:

\begin{equation}
u\cot\theta \partial_u \Psi_1^{4,0} +\partial_\theta \Psi_1^{4,0} +3\cot\theta \Psi_1^{4,0}=3u\Psi_2^{3,0}.
\end{equation}
The constraint equation for the leading coefficients of  $\Psi_0^{5}$ for a non-polyhomogeneous spacetime ($\Psi_0^{5,0}$), and a ``minimal'' polyhomogeneous spacetime ($\Psi_0^{5,1}$) are respectively,
\begin{equation}
u \cot\theta \partial_u \Psi_0^{5,0} +\partial_\theta \Psi_0^{5,0}+3\cot\theta \Psi_0^{5,0}=2u\Psi_1^{4,0},
\end{equation}
and
\begin{equation}
u \cot\theta \partial_u \Psi_0^{5,1} +\partial_\theta \Psi_0^{5,1}+3\cot\theta \Psi_0^{5,1}=0.
\end{equation}
The homogeneous part of the solutions of all these equations will be of the form $u^{-3}F(u^{-1}\sin \theta)$.

Note that in the ``minimal'' polyhomogeneous case, no source/sink term occurs. The solution in this case is therefore:
\begin{equation}
\Psi _0^{5,1}=\frac 1{u^3}F_2\left( \frac{\sin \theta }u\right), 
\end{equation}
where $F_2$ is an arbitrary function of the argument. Now, using the evolution equation for $\Psi_0^{5,1}$ (see \cite{javk98a} or \cite{javk98c}), one finds that
\begin{equation}
\dot{\Psi}_0^{5,1}=\eth_1\Psi_1^{4,1}=0,
\end{equation}
whence $\Psi_0^{5,1}$ will also be a constant of motion for boost-rotation symmetric spacetimes. Hence
\begin{equation}
\Psi_0^{5,1}=C_2 \csc^3 \theta.
\end{equation}
Again one can see that polyhomogeneity in these class of spacetimes gives rise to a local $\scri$.

\subsection{Constraints on $\Psi _0^{6}$ and the NP constants.}

Finally, we are able to  deduce constraint equations for $\Psi _0^{6}$. As seen in section 2.3, the logarithmic NP constants are given in terms of an integral of $\Psi_0^{6,X}$, where $X=N_5$ if $N_3=-\infty$. The constraint equation will be much more complicated in this case. The expansion at order 6 in $1/r$ of the integrability condition (I0) (equation (\ref{I0})) gives an equation of the form:

\begin{equation}
b\dot{\Psi}_0^6-c_{-1}\eth_1\Psi_0^6-\cbar_{-1}\ethbar_1\Psi_0^6 -2(a_{-1}+\eth_1c_{-1})\Psi_0^6 = K,
\end{equation}
where $K$ is a complicated expression depending on $\Psi_0^4$, $\Psi_0^5$ and their $\eth_1$ and $\ethbar_1$ derivatives, and on $\Psi_1^{4,0}$, $\Psi_2^{3,0}$, $a_{-1}$, $a_0$, $a_1$, $b$, \& $c_{-1}$, $c_0$, $c_1$. In the particular case of the non-polyhomogeneous spacetime the coefficient which yields the Newman-Penrose constants $\Psi_0^{6,0}$ satisfies,

\begin{eqnarray}
&b\dot{\Psi}_0^{6,0}-c_{-1}\eth_1\Psi_0^{6,0}-\cbar_{-1}\ethbar_1\Psi_0^{6,0} -2(a_{-1}+\eth_1c_{-1})\Psi_0^{6,0} =&\nonumber  \\
&(\frac{5}{2}b +8\cbar_{-1}\eth_1\sigmab_{2,0}+6c_{-1}\ethbar_1\sigma_{2,0}-2\ethbar_1c_{-1}\sigma_{2,0}+2\eth_1\cbar_{-1}\sigmab_{2,0})\Psi_0^{5,0}& \nonumber \\
&-c_{-1}\sigma_{2,0}\ethbar_1\Psi_0^{5,0}+c_0\eth_1\Psi_0^{5,0}-\cbar_{-1}\sigmab_{2,0}\eth_1\Psi_0^{5,0}+\cbar_0\ethbar_1\Psi_0^{5,0}& \nonumber \\
&-4\eth_1b\Psi_1^{5,0}+4\sigma_{2,0}\ethbar_1b\Psi_1^{4,0}-4b\tau_2\Psi_1^{4,0},& \label{psi060}
\end{eqnarray}
and the analogous coefficient ($\Psi_0^{6,1}$) for a ``minimal'' polyhomogeneous spacetime has to satisfy a similar constraint equation

\begin{eqnarray}
&b\dot{\Psi}_0^{6,1}-c_{-1}\eth_1\Psi_0^{6,1}-\cbar_{-1}\ethbar_1\Psi_0^{6,1} -2(a_{-1}+\eth_1c_{-1})\Psi_0^{6,1} =&  \nonumber \\
&(\frac{5}{2}b +8\cbar_{-1}\eth_1\sigmab_{2,0}+6c_{-1}\ethbar_1\sigma_{2,0}-2\ethbar_1c_{-1}\sigma_{2,0}+2\eth_1\cbar_{-1}\sigmab_{2,0})\Psi_0^{5,1}& \nonumber \\
&-c_{-1}\sigma_{2,0}\ethbar_1\Psi_0^{5,1}+c_0\eth_1\Psi_0^{5,1}-\cbar_{-1}\sigmab_{2,0}\eth_1\Psi_0^{5,1}+\cbar_0\ethbar_1\Psi_0^{5,1}& \nonumber \\
&+4\eth_1b\ethbar_1\Psi_0^{5,1}+4\sigma_{2,0}\ethbar_1b\ethbar_1\Psi_0^{4,0}-4b\tau_2\ethbar_1\Psi_0^{4,0},& \label{psi061}
\end{eqnarray}
The source/sink terms in these two equations are much more complicated than those in previous sections, and hence their interpretation is not that clear cut.

Using the ideas and notation of section 4.1, one can split $\Psi_0^{6,X}$ into its homogeneous $(\Psi_0^{6,X})_h$ and non-homogeneous parts $(\Psi_0^{6,X})_p$. From equations (\ref{xh}) and (\ref{xp}) one finds that (as path of the line integrals reduces to a point)

\begin{equation}
(\Psi_0^{6,X})_h|_{\N_0}= f(\theta,\varphi),
\end{equation}
and that

\begin{equation}
(\Psi_0^{6,X})_p|_{\N_0}= 0.
\end{equation}
The Newman-Penrose constants can be evaluated on any null hypersurface, in particular at $\N_0$, therefore

\begin{equation}
\mathcal{Q}_k^X= \int_{S^2} f(\theta,\varphi) (_2Y_{2,k}) \d S.
\end{equation}
The conservation of $\mathcal{Q}_k^X$ shows that the added effect over time of the complicated source terms of equations (\ref{psi060}) and (\ref{psi061}) cancels exactly the damping of the initial data $f(\theta,\phi)$.

\subsubsection{ Boost rotation symmetry.}

The constraint equations for $\Psi_0^{6,0}$ and $\Psi_0^{6,1}$ in the case of boost-rotation symmetry spacetimes are of the form: 
\begin{equation}
u\cot \theta \partial _u\left( \Psi _0^{6,0}\right) +\partial _\theta
\left( \Psi _0^{6,0}\right) +4\cot \theta \left( \Psi _0^{6,0}\right)=H_X.
\end{equation}

For the ``minimal'' polyhomogeneous spacetime the non-homogeneous term simplifies to (recall that $\sigma_{2,0}=u^{-2}F_1(u^{-1}\sin\theta)$ and $\Psi_0^{5,1}=C_2\csc^3\theta$)
\begin{equation}
H_1=-C_2 \csc^3(\theta)\cot(\theta) u^{-2}F_1\left(\frac{\sin\theta}{u}\right).
\end{equation}
Finally, the homogeneous part of the solutions to the constraint equations have the form
\begin{equation}
\frac{1}{u^4}F_3\left(\frac{\sin\theta}{u}\right).
\end{equation}

\section{Conclusions}
We have seen that if an asymptotically flat spacetime ---polyhomogeneous or not--- is assumed to have a Killing vector field, then using the integrability conditions of the Killing equations it is possible to obtain constraints on the different components of the Weyl tensor and the news function. These extra equations together with the evolution equations derived from the Bianchi identities suggest the existence, in non-stationary spacetimes, of processes of transformation of multipole moments of a given kind into others of a different class. Unfortunately, this interpretation is done in terms of quantities that can only be defined rigorously for stationary spacetimes.

Newman \& Penrose \cite{np68} found that for a stationary spacetime the Newman-Penrose constants have the structure

\begin{equation}
\mbox{(dipole)}^2-\mbox{(monopole)}\times\mbox{(quadrupole)}.
\end{equation}
If one considers a system that initially is stationary, and later undergoes a process of gravitational radiation, finally settling down into a stationary state, one can see that the NP constants impose a ``selection rule'' for the class of final states achievable. The idea of transformation of multipole moments discussed above fits with this idea, as the monopole moment of the source will be radiated according to Bondi's mass loss formula, the dipole and quadrupole moments changing accordingly in order to preserve the value of the NP constants.

\section*{Acknowledgements}
I am most grateful to my supervisor Prof M A H MacCallum for numerous discussions, advice and encouragement. I thank Prof J Bi\v{c}\'{a}k for inviting me to give a seminar in Prague, for his warm hospitality, and for giving me references, and to Dr A Pravdov\'{a} for acquainting me with their work. I also thank Dr G Kerr for several discussions on spin-weighted quantities, and Killing equations in the GHP/NP formalism, and t o Dr R Lazkoz for checking some calculations, helping with the graphs and making useful suggestions on the draft. I hold a scholarship (110441/110491) from the Consejo Nacional de Ciencia y Tecnolog\'{\i}a (CONACYT), Mexico.

\appendix

\section{The solution to the equation: $\alpha =c_{-1}\eth_1 \beta +\overline{c%
}_{-1}\ethbar_1  \beta $.}

The function $\beta $ is of spin weight zero, hence

\begin{equation}
\eth_1 \beta =\frac 1{\sqrt{2}}\left\{ \frac \partial {\partial \theta }-\frac
i{\sin \theta }\frac \partial {\partial \varphi }\right\} \beta .
\end{equation}

Write $c_{-1}=c_r+ic_i$. Then we see that

\begin{equation}
c_{-1}\eth_1 \beta +\cbar_{-1}\ethbar_1 \beta =\sqrt{2}\left( c_r%
\frac{\partial \beta }{\partial \theta }+\frac{c_i}{\sin \theta }\frac{%
\partial \beta }{\partial \varphi }\right) ,
\end{equation}

hence the partial differential equation to be solved is

\begin{equation}
\alpha =\sqrt{2}\left( c_r\frac{\partial \beta }{\partial \theta }+\frac{c_i%
}{\sin \theta }\frac{\partial \beta }{\partial \varphi }\right) .
\label{betaeqn}
\end{equation}

Recall that both $\alpha $ and $\beta $ can be expanded in terms of
spherical harmonics as

\begin{eqnarray}
\alpha &=&\sum_l\sum_m\alpha _{lm}Y_{lm}, \\
\beta &=&\sum_l\sum_m\beta _{lm}Y_{lm}.
\end{eqnarray}

Now, the right-hand side of equation (\ref{betaeqn}) can also be expanded
in terms of spin zero spherical harmonics as 

\begin{equation}
\sum_L\sum_Mf_{LM}(A_m,\beta _{lm})Y_{LM},
\end{equation}
where $f_{LM}$ is linear in the $\beta _{lm}$. This gives rise to the
following infinite set of linear equations in $\beta _{lm}$

\begin{equation}
\alpha _{LM}=f_{LM}(A_m,\beta _{lm}),
\end{equation}
that in principle can be solved to any desired order, yielding a solution to
the equation (\ref{betaeqn}).

\section{The solution to the equation: $u\cot \theta \partial _ux+\partial
_\theta x+k\cot \theta x=H(u,\theta )$.}

It is not very complicated to find the solution to the initial value problem

\begin{equation}
u\cot \theta \frac{\partial x}{\partial u}+\frac{\partial x}{\partial \theta 
}+k\cot \theta x=H(u,\theta ),  \label{continuity}
\end{equation}
with initial data

\begin{equation}
x(1,\theta )=F(\theta ).
\end{equation}
The associated system of ordinary differential equations is

\begin{eqnarray}
\frac{du}{d\eta } &=&u\cot \theta ,  \label{eqn1} \\
\frac{d\theta }{d\eta } &=&1,  \label{eqn2} \\
\frac{dx}{d\eta } &=&-k\cot x+H(u,\theta ),  \label{eqn3}
\end{eqnarray}
with initial data

\begin{eqnarray}
u(0) &=&1, \\
\theta (0) &=&\xi , \\
x(0) &=&F(\xi ).
\end{eqnarray}
Equation (\ref{eqn2}) gives

\begin{equation}
\theta =\xi +\eta ,
\end{equation}
and so we can solve now equation (\ref{eqn1}) yielding

\begin{equation}
u=\frac 1{\sin \xi }\sin (\eta +\xi ).
\end{equation}
Therefore the characteristics of the differential equation are given by

\begin{equation}
u=\frac 1{\sin \xi }\sin \theta.
\end{equation}
It will be necessary to invert this last expression in order to have $\xi$ as a function of $u$ and $\theta$. The range of $\theta$ is in principle $[0,\pi)$. There will be problems with the invertibility whenever $u=\sin \theta$. The solution will break down there. This comes from the fact that the polar coordinates $(\theta,\varphi)$ are not good coordinates for the sphere, and several coordinate patches are needed to cover it.

Equation (\ref{eqn3}) is more involved,

\begin{equation}
\frac{dx}{d\eta }=-k\cot (\xi +\eta )x+H\left( \frac 1{\sin \xi }\sin (\eta
+\xi ),\xi +\eta \right).
\end{equation}
The solution to the homogeneous equation can be obtained directly by
integration

\begin{equation}
x_h=\frac 1{u^k}F\left( \frac{\sin \theta }u\right) ,
\end{equation}
while the particular solution to the non-homogeneous equation is given in an
integral form

\begin{equation}
x_p=\csc ^k\left( \eta +\xi \right) \int_0^\eta H\left( \frac{\sin \left(
t+\xi \right) }{\sin \xi },t+\xi \right) \sin ^k\left( t+\xi \right) \d t
.
\end{equation}

Then the solution to the differential equation is

\begin{eqnarray}
x(u,\theta ) &=&\frac 1{u^k}F\left( \frac{\sin \theta }u\right)  \nonumber
\label{generalsol} \\
&&+\csc ^k\left( \theta \right) \left[ \int_0^{\theta -\xi }H\left( \frac{%
\sin \left( t+\xi \right) }{\sin \xi },t+\xi \right) \sin ^k\left( t+\xi
\right) \d t\right] _{\xi =\arcsin \left( \sin \theta /u\right) }.
\label{gensol}
\end{eqnarray}

The key to interpreting the solution (\ref{gensol}) is to regard equation (\ref
{continuity}) as a continuity equation completely analogous to the $\rho
,_t+\nabla \cdot J=Source$ continuity equation of fluid mechanics. The terms 
$u\cot \theta \partial _ux+\partial _\theta x$ correspond to the time
derivative plus divergence bit of the continuity
equation, while the $k\cot \theta x$ term is a damping term due to the
interaction of the gravitational field with itself (note that the constant $%
k $ gives rise in the solution to a $1/u^k$ term that diminishes the
amplitude of the initial data. Finally, the function $H$ works like a
source/sink term.

\section{The orbits on the sphere of the Killing vector fields}

As seen in section 4, the restriction of the Killing vector field to $\scri$ is given by

\begin{equation}
\xi^\mu|_\scri=\left( -a_{-1}u+\alpha, -\frac{1}{\sqrt{2}}(\cbar_{-1}+c_{-1}), -\frac{i\csc \theta}{\sqrt{2}}(\cbar_{-1}-c_{-1}) \right).
\end{equation}
The integral curves of this vector field in $\scri$ can be conveniently visualized as curves on the unit sphere parametrized by the retarded time $u$. The structure of this Killing vector field is not very complicated, depending only on 3 spherical harmonics. Therefore, it is quite tempting to attempt a study of these orbits.

Throughout this article, we have been using the angular coordinates $(\theta,\varphi)$ on the cuts of $\scri$, however the study of the orbits is more easily done using the stereographic coordinates ($\zeta=\cot \frac 12 \theta e^{i\theta}$), and working on the complex plane plus the point at infinity. The ${_1Y_{1,m}}$ spherical harmonics in stereographic coordinates are given by \cite{stewart}:

\begin{eqnarray}
{_1Y_{1,1}}&=& -\sqrt{\frac{3}{4\pi}} \frac{\zeta^2}{1+\zeta\zetab} \\
{_1Y_{1,0}}&=& -\sqrt{\frac{3}{2\pi}} \frac{\zeta}{1+\zeta\zetab}  \\
{_1Y_{1,-1}}&+& -\sqrt{\frac{3}{4\pi}} \frac{1}{1+\zeta\zetab}
\end{eqnarray}
Now, the critical points of the vector field correspond to the points on the sphere (complex plane) where $c_{-1}=0$. Using the spin-weighted spherical harmonics in spherical coordinates one can prove the following lemma.

\begin{lemma}
The restriction of a Killing vector field of an asymptotically flat spacetime to $\scri$ vanishes at most in two points of the sphere (and at least in one).
\end{lemma}

The result follows noting that if  $c_{-1}=0$, then one has

\begin{equation}
A_{1}\zeta^2 + A_{0} \zeta + A_{-1} =0,
\end{equation}
a second degree equation in \Complex. From the Fundamental Theorem of Algebra, one knows that the equation has two roots.

The vector field will vanish at least in one point, so without loss of generality one can set this point to be the origin of the complex plane (the North Pole of the sphere), and the other root (if present) to lie on the real axis. Hence one can write

\begin{equation}
\cbar_{-1}= e^{i\omega}\frac{a\zeta^2 +b\zeta}{1+\zeta\zetab}.
\end{equation}

Let $\zeta=x+iy$, then

\begin{eqnarray}
\mbox{Re} \, \cbar_{-1}= \frac{(a(x^2-y^2)+bx)\cos\omega -(2axy+by)\sin\omega}{1+x^2+y^2}, \\
\mbox{Im} \, \cbar_{-1}= \frac{(a(x^2-y^2)+bx)\sin\omega +(2axy+by)\cos\omega}{1+x^2+y^2}.
\end{eqnarray}
And the orbits on the complex plane are given by

\begin{equation}
\frac{\d y}{\d x}= \frac{\mbox{Im}\, \cbar_{-1}}{\mbox{Re} \, \cbar_{-1}}= \frac{(a(x^2-y^2)+bx)\sin\omega +(2axy+by)\cos\omega}{(a(x^2-y^2)+bx)\cos\omega -(2axy+by)\sin\omega}. \label{orbits}
\end{equation}

There is no integrating factor for this equation for an arbitrary $\omega$, however the form of the orbits can be found readily by noting that they will have horizontal tangent at the ``rotated'' hyperbola

\begin{equation}
a\sin\omega x^2+2a\cos\omega xy -a\sin \omega y^2 + b\sin\omega +b\cos\omega y=0,
\end{equation}
and vertical tangency at another hyperbola

\begin{equation}
a\cos\omega x^2 -2a\sin\omega xy -a\cos\omega y^2 + b\cos\omega x -b\sin\omega y=0.
\end{equation}
These two curves intersect by construction only at $(0,0)$ and $(-b/a,0)$ (the critical points). If $\omega \not =0$, $\pi/2$, $\pi$, or $3\pi/2$ then none of the hyperbolae are degenerate (i.e. they are not intersecting lines). The degenerate cases will be studied separately. Linearizing the differential equation (\ref{orbits}) around the origin one obtains (if $b\not=0$) 

\begin{equation}
\frac{ \d y}{\d x}= \frac{ \sin\omega x + \cos\omega y}{\cos\omega x - \sin\omega y},
\end{equation}
The solution of this equation is  the logarithmic spiral

\begin{equation}
x^2+y^2= C e^{2\cot\omega \arctan(y/x)}. \label{spiral}
\end{equation}
A similar result follows when linearizing around $(-b/a,0)$. Therefore if $a$ and $b$ are different from zero and $\omega \not =0$, $\pi/2$, $\pi$, or $3\pi/2$ then the orbits spiral around $(0,0)$ and $(-b/a,0)$.

\begin{figure}[hbt]
\centering
  \begin{tabular}{c@{\qquad}c}
  \mbox{ \epsfig{file=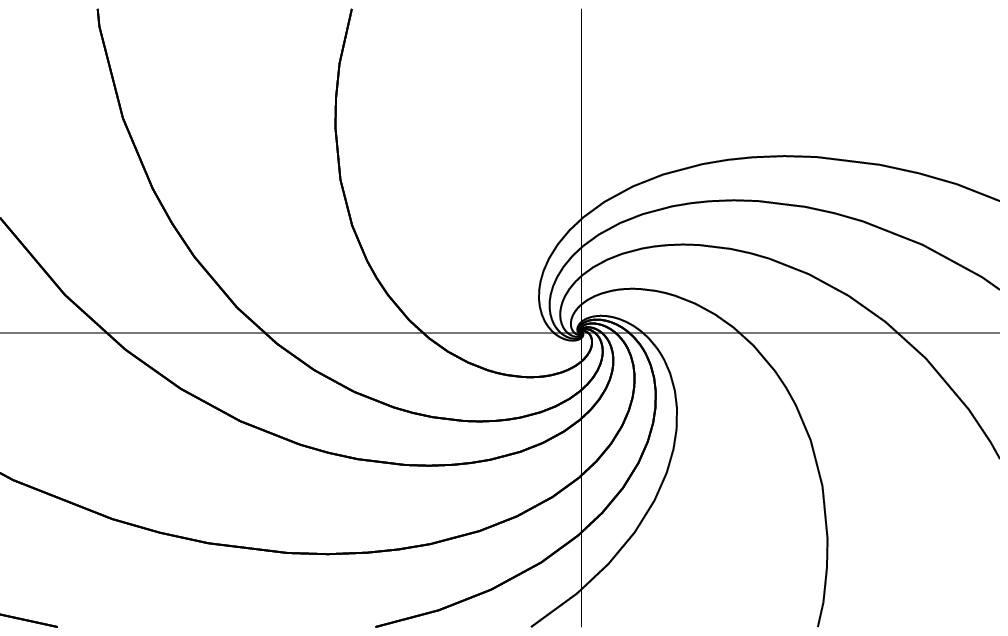, width=5cm, height=5cm}} &
  \mbox{ \epsfig{file=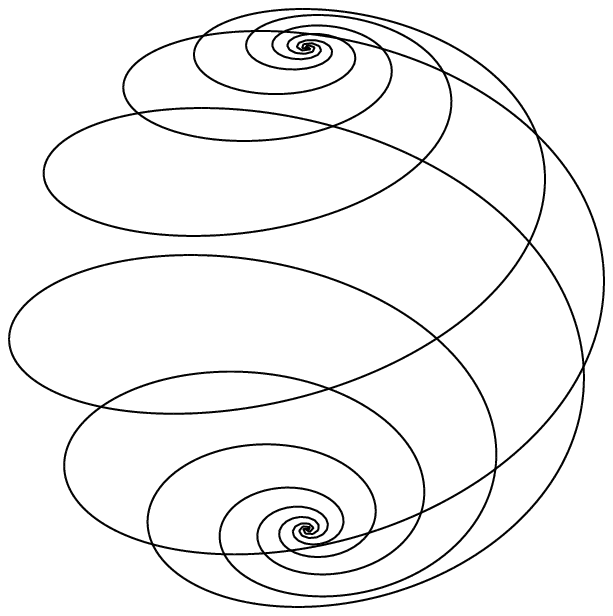, width=5cm, height=5cm}}
  \end{tabular}
\caption{Spiral orbits on the complex plane and on the sphere.}
\end{figure}

In the case that  $\omega=0$, $\pi$ then the differential equation for the orbits (\ref{orbits}) can be readily solved yielding

\begin{equation}
\left( x+ \frac{b}{2a} \right)^2 + \left(y-\frac{C}{2a} \right)^2 = \left( \frac{C}{2a} \right)^2 + \left( \frac{b}{2a} \right)^2.
\end{equation}
That is, the orbits are circles that intersect each other in the points $(0,0)$ and $(-b/a,0)$. 

\begin{figure}[hbt]
\centering
  \begin{tabular}{c@{\qquad}c}
  \mbox{ \epsfig{file=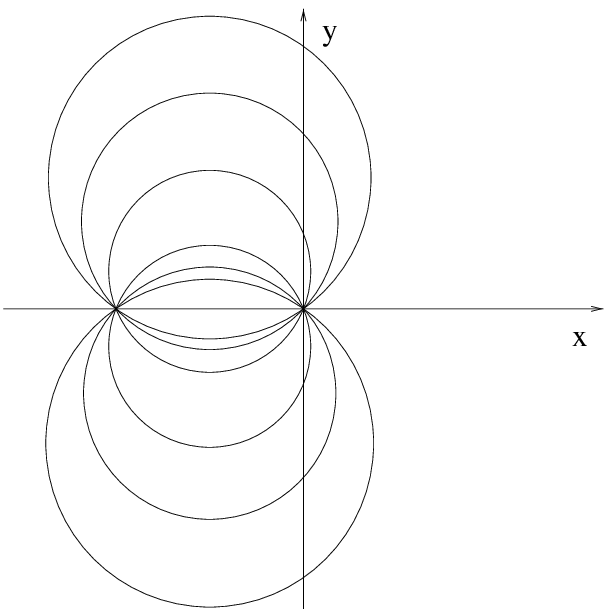, width=5cm, height=5cm}} &
  \mbox{ \epsfig{file=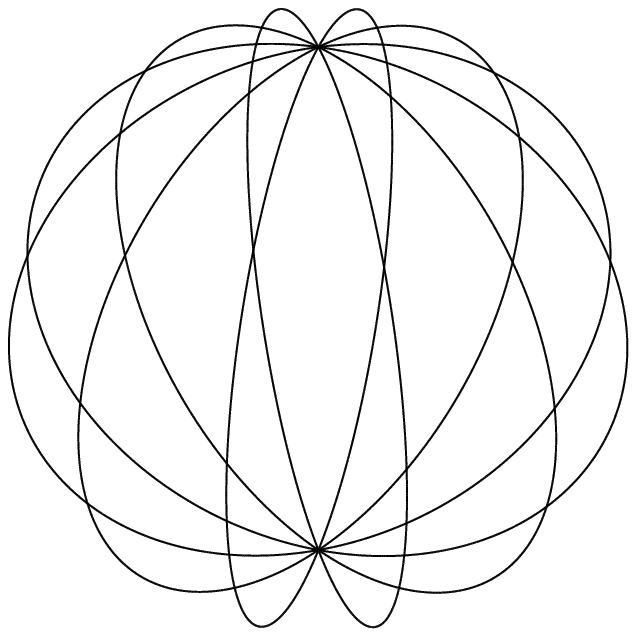, width=5cm, height=5cm}}
  \end{tabular}
\caption{``Electric dipole''-like orbits on the complex plane and on the sphere}
\end{figure}

Now, if $a$, $b\not=0$ and $\omega=\pi/2$, $3\pi/2$ then the orbits are non-intersecting circles whose centres lie on the x axis, and have a reflexion symmetry along the line $x=-b/2a$.

If $b=0$ then the two critical points coincide, and one gets a family of tangent circles at the origin. The angle $\omega$ yields the orientation of the family. 

If $a=0$, then one of the critical points is at the origin (North Pole), and the other at infinity (South Pole). The generic orbit ($\omega\not=0$, $\pi/2$, $\pi$, $3\pi/2$) is the logarithmic spiral described by equation (\ref{spiral}). If $\omega=0$, $\pi$ then one obtains lines that intersect at the origin and at infinity,

\begin{equation}
y=Cx.
\end{equation}
while if $\omega=\pi/2$, $3\pi/2$ then the orbits are concentric circles.
\begin{figure}[hbt]
\centering
 \begin{tabular}{c@{\qquad}c}
  \mbox{ \epsfig{file=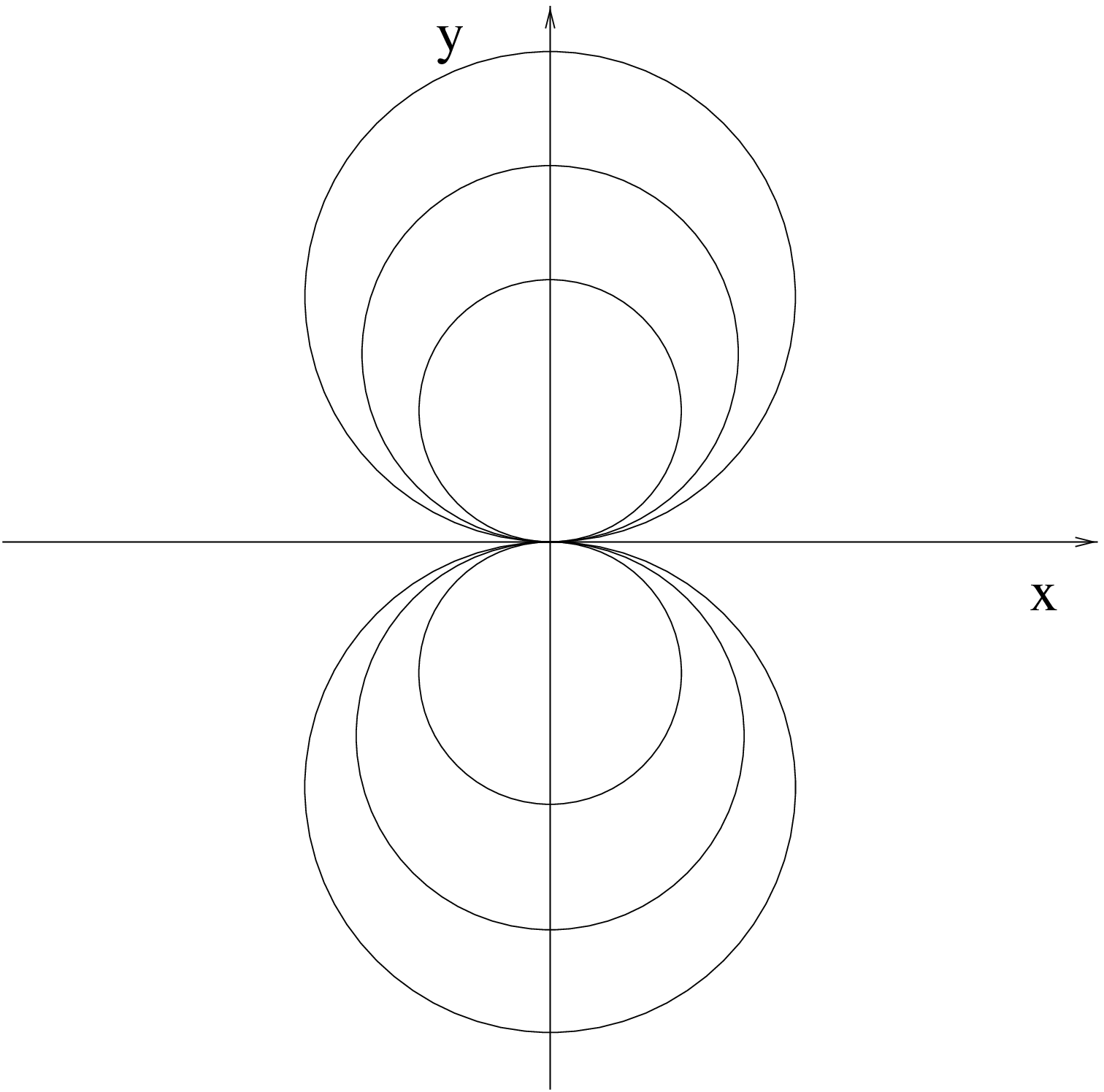, width=5cm, height=5cm}} &
  \mbox{ \epsfig{file=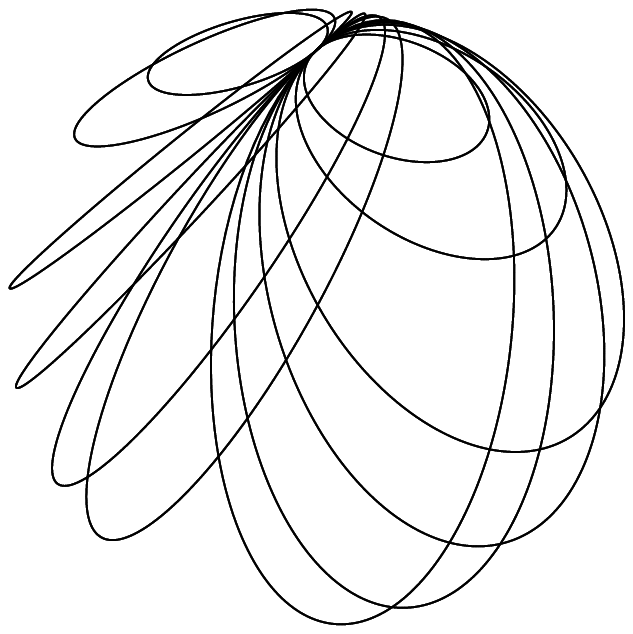, width=5cm, height=5cm}}
  \end{tabular}
\caption{Orbits with a single point on the complex plane and on the sphere.}
\end{figure}

\begin{figure}[hbt]
\centering
  \begin{tabular}{c@{\qquad}c}
  \mbox{ \epsfig{file=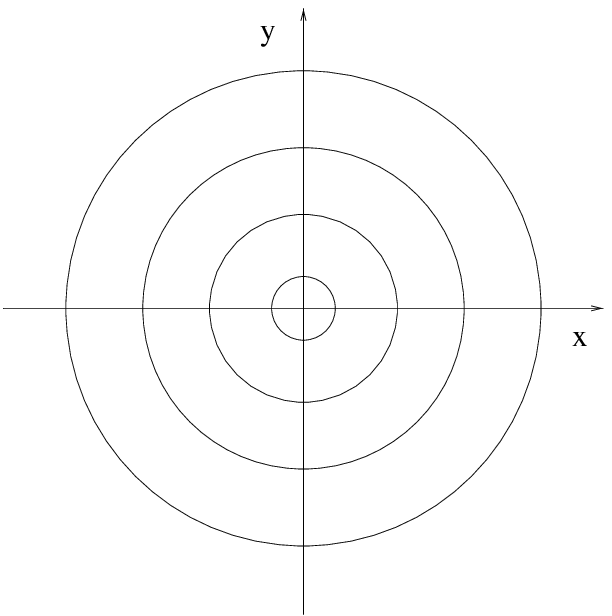, width=5cm, height=5cm}} &
  \mbox{ \epsfig{file=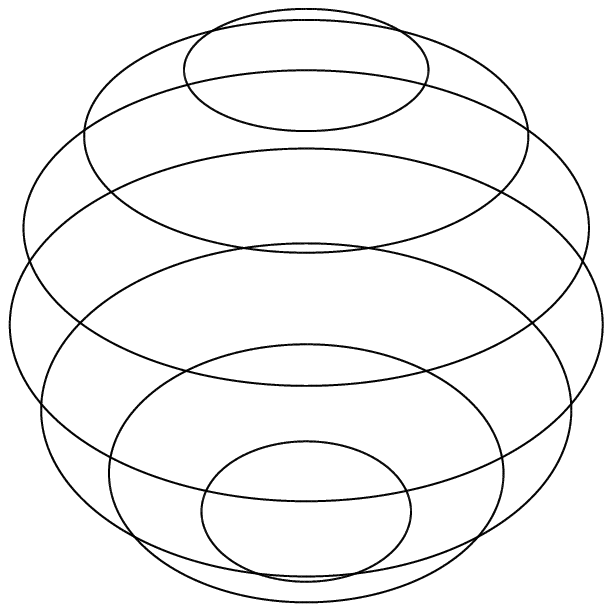, width=5cm, height=5cm}}
  \end{tabular}
\caption{Periodic orbits on the complex plane and on the sphere.}
\end{figure}

The previous discussion gives a complete classification of asymptotically flat spacetimes with one Killing vector field in terms of their orbits at $\scri$. The results are collected in the following proposition:

\begin{proposition}
Let \cal{M} be an asymptotically flat spacetime with complete $\scri$ and one non supertranslational Killing vector field. Then the orbits of the restriction of the Killing field to $\scri$ ($\xi^\mu|_\scri$) regarded as curves on $S^2$ parametrised by $u$ satisfy one of the following:

\begin{enumerate}
 \item there are two critical points, and the orbits spiral around them;
 \item there are two critical points, and the orbits connect the two points (``electric dipole'' structure);
 \item there are two critical points, and the orbits are periodic around them;
 \item there is one critical point, and the orbits are closed curves tangent at them.
\end{enumerate}
\end{proposition}
Observe that the critical points are not necessarily antipodes (i.e. located at opposite points on the sphere).


\section{Some spherical harmonics}

\subsubsection{Spin-weight 0}
\begin{eqnarray}
{_0 Y_{0,0}}&=& \frac{1}{\sqrt{4\pi}} \\
{_0 Y_{1,1}}&= &-\sqrt{\frac{3}{8\pi}}\sin\theta e^{i\phi} \\
{_0 Y_{1,0}}&=& \sqrt{\frac{3}{4\pi}} \cos\theta \\
{_0 Y_{1,-1}}&=& \sqrt{\frac{3}{8\pi}} \sin\theta e^{-i\phi}
\end{eqnarray}

\subsubsection{Spin-weight 1}
\begin{eqnarray}
{_1 Y_{1,1}}&=& \sqrt{\frac{3}{16\pi}} (\cos\theta +1) e^{i\phi} \\
{_1 Y_{1,0}}&=& \sqrt{\frac{3}{8\pi}} \sin \theta \\
{_1 Y_{1,-1}} &=& -\sqrt{\frac{3}{16\pi}} (\cos\theta -1) e^{-i\phi}
\end{eqnarray}

\subsubsection{Spin-weight 2}
\begin{eqnarray}
{_2 Y_{2,2}}&=& 3\sqrt{\frac{5}{96\pi}}(1+\cos\theta)^2 e^{2i\phi}  \\
{_2 Y_{2,1}}&=& 3\sqrt{\frac{5}{24\pi}}\sin\theta(1+\cos\theta) e^{i\phi}  \\
{_2 Y_{2,0}}&=&  \frac{3}{2}\sqrt{\frac{5}{4\pi}} \sin^2\theta \\
{_2 Y_{2,-1}}&=& 3\sqrt{\frac{5}{24\pi}}\sin\theta(1-\cos\theta) e^{-i\phi} \\
{_2 Y_{2,-2}}&=& 3\sqrt{\frac{5}{96\pi}}(1-\cos\theta)^2 e^{-2i\phi}
\end{eqnarray}

\end{document}